\documentclass[12pt]{article}

\usepackage{amsmath}
\usepackage{amsfonts}
\usepackage{amssymb}
\usepackage{latexsym}
\usepackage{ifthen}
\usepackage{theorem}
\usepackage{graphicx}

\theorembodyfont{\rmfamily}
\newtheorem{prop}{Proposition}[section]
\newtheorem{lemma}[prop]{Lemma}
\newtheorem{theorem}[prop]{Theorem}
\newtheorem{cor}[prop]{Corollary}

\newcommand{\R}{{\mathbb R}}
\newcommand{\ra}{\rightarrow}

\DeclareMathOperator*{\supp}{supp}

\begin{document}

\title{An algorithm for computing cutpoints in finite metric spaces}

{\tiny
\author{\bf{Andreas Dress}\\
\small{CAS-MPG Partner Institute and Key Lab for Computational Biology (PICB)}\\
\small{Shanghai Institutes for Biological Sciences (SIBS)}\\
\small{Chinese Academy of Sciences (CAS)}\\
\small{320 Yue Yang Road, Shanghai 200031, P.\,R.\,China}\\
\small{email: andreas@picb.ac.cn}\\
\bf{Katharina T. Huber}\\
\small{University of East Anglia, School of Computing Sciences}\\
\small{Norwich, NR4 7TJ, UK}\\
\small{e-mail: Katharina.Huber@cmp.uea.ac.uk}\\ 
\bf{Jacobus Koolen}\\
\small{Department of Mathematics and Pohang Mathematics Institute}\\
\small{POSTECH}\\
\small{Pohang, South Korea}\\
\small{email: koolen@postech.ac.kr}\\ 
\bf{Vincent Moulton}\\
\small{University of East Anglia, School of Computing Sciences}\\
\small{Norwich, NR4 7TJ, UK}\\
\small{e-mail: vincent.moulton@cmp.uea.ac.uk}\\ 
\bf{Andreas Spillner}\\
\small{University of Greifswald, Department of Mathematics and Computer Science}\\
\small{17489 Greifswald, Germany}\\
\small{e-mail: andreas.spillner@uni-greifswald.de}\\
}}

\date{}

\maketitle

\newpage 

\begin{abstract}
The theory of the {\em tight span}, a  
cell complex 
that can be associated to every metric \(D\), 
offers a unifying view on existing approaches
for analyzing distance data, in particular 
for decomposing a
metric \(D\) 
into a sum of simpler metrics
as well as for representing it
by certain specific edge-weighted graphs,
often referred to as 
{\em realizations} of \(D\).
Many of these approaches involve the explicit or implicit
computation of the so-called cutpoints of 
(the tight span of) \(D\),
such as the algorithm for computing the
``building blocks'' of
optimal realizations of $D$ recently presented by A.Hertz and S.Varone.
The main result of this paper is an algorithm for computing
the set of these cutpoints for a metric \(D\) on a finite
set with \(n\) elements in $O(n^3)$ time.
As a direct consequence, this improves the run time of the aforementioned
$O(n^6)$-algorithm by Hertz and Varone by ``three orders of magnitude''.
\end{abstract}


%%%%%%%%%%%%%%%%%%%%%%%%%%%%%%%%%%%%%%%%%%%%%%%%%%%%%%%%%%%%%%%%%%%%%%%%
\section{Introduction}
\label{section:introduction}
%%%%%%%%%%%%%%%%%%%%%%%%%%%%%%%%%%%%%%%%%%%%%%%%%%%%%%%%%%%%%%%%%%%%%%%%

The decomposition of a given metric into simpler metrics 
(see e.g. \cite{deza:laurent:cuts:1997})
is a fundamental problem in classification featuring  applications in, for example, 
clustering (e.g. \cite{bryant:berry:clustering:2001}), 
``networking'' (e.g. \cite{chu-gar-gra-01a}), and 
phylogenetics (e.g. \cite{huson:bryant:networks2005}). 
The theory of the \emph{tight span} \[
T(D) := \{f \in \mathbb{R}^X :
f(x) = \underset{y \in X}{\sup} \big(D(x,y)-f(y)\big) \ 
\text{for all} \ x \in X\},
\]
of a metric \(D\) defined on a set \(X\) \cite{isbell:six:theorems:1964,dress:tight:extensions:1984}
offers a unifying view on various existing approaches developed for 
this task. 
In this paper, we focus on decompositions of metrics \(D\) 
defined on a finite set $X$ that are induced by \emph{cutpoints} of \(T(D)\), that is, maps \(f \in T(D)\) 
such that \(T(D)-\{f\}\) is disconnected. These decompositions
are closely related to  certain \emph{graph realizations} of \(D\),
that is, connected edge-weighted graphs \(G = (V,E,\ell: E \rightarrow \mathbb{R}_{>0})\) with \(X \subseteq V\) 
for which \(D(x,y) = D_G(x,y)\) holds for all \(x,y \in X\) (where \(D_G\) denotes the shortest-path metric induced by \(G\) on \(V\)).

\begin{figure}
\centering
\includegraphics[scale=1.0]{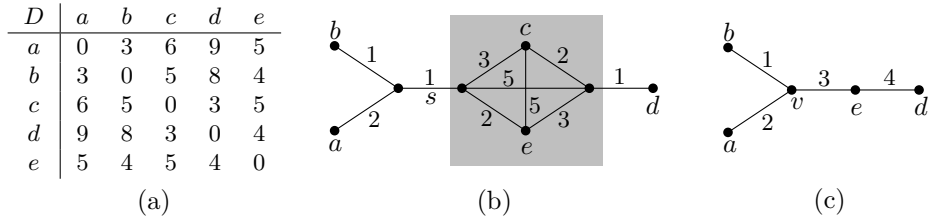}
\caption{(a) An example of a metric \(D\) on \(X=\{a,\dots,e\}\).
         (b) A block realization of \(D\): The vertices in the shaded
             region form a block and edge \(s\) is a bridge.
         (c) A block realization of the restriction \(D'\) of \(D\) to
             the subset \(X' := X \setminus \{c\}\).}
\label{figure:block:realization}
\end{figure}

To describe these graph realizations,
recall (see e.g.\,\cite{wes-96a})
that a vertex \(v\) in a  
graph \(G = (V,E)\) is called a 
\emph{cut vertex} (of $G$) if there exist vertices $u,u'\in V$ with $\{u,v\}, \{u',v\}\in E$  such that every path from $u$ to $u'$ in $G$ passes through $v$. 
Moreover, a maximal subset \(B \subseteq V\) 
with the property that the induced graph \(G[B] := (B, E \cap \binom{B}{2})\)
has no cut vertex is called a \emph{block} of \(G\).
A graph realization \(G\) of \(D\) is called a \emph{block realization} of \(D\) if \(G\) is a {\em block graph}, i.e., every 
block of \(G\) is a clique, and every vertex in \(V \setminus X\) has degree at least 3 and
is a cut vertex of \(G\). 
An example of a block realization is presented in Figure~\ref{figure:block:realization}(b).

In a recent series of papers 
\cite{dre-hub-koo-08a,dress:huber:compatible:decompositions:2008,dress:huber:koolen:moulton:cut:points:2007},
it has been observed 
that a map \(f \in T(D)\) is a  cutpoint if and only
if the graph
$\Gamma_f:=(X_f,E_f)$ defined, for every \(f \in \mathbb{R}^X\), by 
$
X_f := \supp(f) $ and  $E_f:=\big\{\{x,y\} \in \binom{\supp(f)}{2}:
f(x) + f(y) > D(x,y)\big\}$ is disconnected (where, as usual, 
\(\supp(f) := \{x \in X : f(x) \neq 0\}\) denotes the \emph{support} of \(f\)), that a map \(f\) in 
\[
P(D) := \{f \in \mathbb{R}^X : f(x) + f(y) \geq D(x,y) \ \text{for all} \ x \in X\}
\]
for which the graph \(\Gamma_f\) is disconnected 
must be contained in --- and, hence, a cutpoint of --- $T(D)$,
and that a cutpoint \(f \in T(D)\) has a neighborhood that is homeomorphic to the open interval 
\((-1,+1)\) if and only if  \(\Gamma_f\) is the disjoint union of two cliques. 
As such maps are of little interest for constructing block realizations, 
we will focus our attention in particular to the set of those cutpoints, denoted by \(cut^*(D)\), 
for which either \(\supp(f) \neq X\) holds or
\(\Gamma_f\) is not the disjoint union of two cliques.

In this paper, we present an algorithm with run time \(O(n^3)\) to compute \(cut^*(D)\),
where \(n = |X|\), improving the run time of the algorithm presented
in \cite{dress:virtual:cutpoints:2007}. Once the set \(cut^*(D)\) is available, 
it is straight-forward to compute a corresponding canonical 
block realization \(G=G_D=(V_D,E_D,\ell_D)\) of \(D\) in \(O(n^3)\) time.
And, from that block realization it is then easy
to derive, for every block $B$ of $G_D$, a 
corresponding metric $D_B$ on $X$ that assigns, to any two elements
$x,x'\in X$, the value $D_B(x,x')$ defined as the
total weight of those edges on any shortest path
from $x$ to $x'$ in $G$ that have both
end points in $B$. For example in Figure~\ref{figure:block:realization}(b) 
the distance $D_B(a,d)$ between
$a$ and $d$ is 5 where $B$ is the block in the
shaded region.

This yields a decomposition of \(D\) into 
a sum of metrics 
of the form 
\(D_B\) where \(B\) runs through the
blocks of \(G_D\) that can
be computed in \(O(n^3)\) time. As a consequence, our algorithm improves 
the run time of the algorithm presented in 
\cite{hertz:varone:cutpoint:partition} 
that follows a 2-step approach: It computes
first those metrics 
\(D_B\) that correspond to blocks \(B\) with only 2 vertices, the so-called
\emph{bridges}, (see \cite{hertz:varone:bridge:partition} for
details) and then the remaining metrics \(D_B\) for the blocks \(B\) that
are not bridges.

Our paper is structured as follows. In the next section, we
introduce the concept of block splits and show how they can help in the computation
of \(cut^*(D)\). In Section~\ref{section:key:properties}, we establish the key properties
of block splits and cutpoints in \(cut^*(D)\) that we use in our new algorithm
for computing \(cut^*(D)\), and
we present this algorithm in Section~\ref{section:algorith:computing:cut:d}.

%%%%%%%%%%%%%%%%%%%%%%%%%%%%%%%%%%%%%%%%%%%%%%%%%%%%%%%%%%%%%%%%%%%%%%%%%%%%%
\section{Block splits}
\label{section:block:splits}
%%%%%%%%%%%%%%%%%%%%%%%%%%%%%%%%%%%%%%%%%%%%%%%%%%%%%%%%%%%%%%%%%%%%%%%%%%%%%

In this section, we present a key concept used
in our algorithm for computing the set \(cut^*(D)\),
where \(D\) is the given metric on a finite set \(X\):
Recall that a \emph{split} \(S\) of \(X\) is a bipartition of \(X\)
into two non-empty subsets \(A\) and \(B\), also denoted by \(A|B\) or \(B|A\).
A split \(A|B\) of \(X\) is called a \emph{block split}
of \(X\) (relative to \(D\)) if there exists a map 
\(f \in P(D)\) with \(\supp(f)=X\)
such that \(\Gamma_f\) is the disjoint union of two cliques whose
vertex sets are \(A\) and \(B\), respectively. 
Note that the condition used in the definition of a block
split above is slightly stronger than the condition
given in \cite[p. 10]{imrich:optimal:realizations:1984}. Also note
that block splits, although not given a specific name,
play an important role in the algorithm for
computing bridges presented in \cite{hertz:varone:bridge:partition}.
The set of block splits of \(X\) induced by \(D\) is  denoted by \(\Sigma_D\). 
In the following, we will also often simply write 
\(xy\) for \(D(x,y)\), \(x,y \in X\).

Our first goal is to establish a property of block splits
that allows to efficiently check whether a given split
of \(X\) is a block split. To this end, we first recall the 
following well-known fact.

\begin{lemma}
\label{lemma:bivariate:map}
Given two sets $A$ and $B$ and a bi-variate map $\phi: A \times B\ra \R$ from the Cartesian product
$A \times B$ into the real numbers (or any Abelian group), there exist maps $\phi_A: A \ra \R$ and $\phi_B: B \ra \R$ with $\phi(a,b)=\phi_A(a)+ \phi_B(b)$ for all $a\in A$ and $b\in B$ if and only if 
$\phi(a,b) + \phi(a',b') = \phi(a,b') + \phi(a',b)$ holds for all $a,a'\in A$ and $b,b'\in B$ if and only if 
$\phi(a,b) + \phi(a_0,b_0) = \phi(a,b_0) + \phi(a_0,b)$ holds for some fixed elements $a_0\in A$ and $b_0\in B$ and all $a\in A$ and $b\in B$.
\end{lemma}

\noindent\textsl{Proof}:
If  there exist maps $\phi_A: A \ra \R$ and $\phi_B: B \ra \R$ with $\phi(a,b)=\phi_A(a)+ \phi_B(b)$, one clearly has $\phi(a,b) + \phi(a',b') = \phi_A(a)+ \phi_B(b)+\phi_A(a')+ \phi_B(b')=
\phi(a,b') + \phi(a',b)$ \ for all $a,a'\in A$ and $b,b'\in B$ while, conversely, if 
$\phi(a,b) + \phi(a_0,b_0) = \phi(a,b_0) + \phi(a_0,b)$ holds for some fixed elements $a_0\in A$ and $b_0\in B$ and all $a\in A$ and $b\in B$, one has 
 $\phi(a,b)=\phi_A(a)+ \phi_B(b)$ for, e.g., the two maps $\phi_A: A \ra \R: a \mapsto \phi(a,b_0)$ and 
$\phi_B: B \ra \R: b \mapsto \phi(a_0,b)-\phi(a_0,b_0)$. 
\hfill\(\blacksquare\)\\

Next, we define, for any map $f\in P(D)$ and any subset $Y$ of $X$, the \emph{virtual distance} $D(f|Y)$ from $f$ to $Y$ (relative to $D$) by
$$ 
D(f|Y):=\frac{1}{2} \min\{f(y)+f(y') - yy': y,y' \in Y\}.$$ 

We will also write $D(x|Y)$ rather that $D(f|Y)$ in case $f$ coincides with the so-called \emph{Kuratowski map} $k_x$ 
\cite{kuratowski:non:separable:metric:spaces:1935} associated with an element $x\in X $ defined by $k_x(y):=xy$ for all $y\in X$.
Note that $0\le D(f|Y) $ holds for all $f$ and $Y$ as above. Note also that, given a split $S=A|B$ of $X$ 
with $ab+a'b'=ab'+a'b$ for all $a,a'\in A$ and $b,b'\in B$, and any two elements $a\in A$ and $b\in B$, one has    
\begin{eqnarray}
\label{definition:isolation:index}
&&D(a|B) + D(b|A) - ab\\
&&\,\,\,\,= 
\frac{1}{2} \ { \min_{ a',a''\in A; b',b'' \in B} }\{ab'+ab'' + a'b + a''b- b'b'' -a'a'' - 2ab \} \notag\\
&&\,\,\,\,= 
\frac{1}{2} \ { \min_{ a',a''\in A; b',b'' \in B} }\{a'b'+a''b''  - b'b'' -a'a''  \} \notag\\
&&\,\,\,\,= \frac{1}{2} \ \min_{a',a'' \in A; b',b''\in B}\{\max(a'b'+a''b'', a'b''+a''b')- a'a''-b'b''\} =: \alpha_S, \notag
\end{eqnarray}
and that $\alpha_S$ has been dubbed the
\emph{isolation index} of $S$ \cite{bandelt:dress:canonical:1992}. 

To illustrate the above definitions, 
note that, 
for the metric given in Figure~\ref{figure:block:realization}(a),
the split $S=\{a,b\}|\{c,d,e\}$ is a block split with 
$D(a|\{c,d,e\})=3$, $D(b|\{c,d,e\})=2$, $D(c|\{a,b\})=4$, $D(d|\{a,b\})=7$, and $D(e|\{a,b\})=3$ and, therefore, 
$D(x|\{c,d,e\})+D(y|\{a,b\})- D(x,y) = 1$, for all $x\in \{a,b\}$ and $y\in \{c,d,e\}$, 
the weight of the edge $s$ separating $\{a,b\}$ from $\{c,d,e\}$ in the corresponding block realization depicted
in Figure~\ref{figure:block:realization}(b).

More generally, we have

\begin{lemma}
\label{lemma:characterization:block:splits}
A split \(S=A|B\) of \(X\) 
is a block split of \(X\) if and only 
if the isolation index $\alpha_{S}$ of $S$ is positive and, choosing
arbitrary elements \(a_0 \in A\) and \(b_0 \in B\),
$a_0b_0+a'b'=a_0b'+a'b_0$ holds for all $a'\in A$ and \(b' \in B\).
\end{lemma}

\noindent\textsl{Proof}:
Assume first that \(S=A|B\) is a block split. By the definition of a block split, 
there exists a map \(f \in P(D)\) for which
\(\Gamma_f\) is the disjoint union of two cliques whose
vertex sets are \(A\) and \(B\) and, therefore, we clearly have 
$D(f|A),D(f|B)>0$. Moreover, for the restrictions $\phi_A:=f|_A$ and 
$\phi_B:=f|_B$ of \(f\) to \(A\) and \(B\), respectively, we have 
\(\phi_A(a) + \phi_B(b) = ab\) for all \(a \in A\) and \(b \in B\), and,
therefore, $ab+a'b'=ab'+a'b$ must hold for all $a,a'\in A$ and
\(b,b' \in B\) in view of Lemma~\ref{lemma:bivariate:map} applied to the bivariate
map \(\phi:=D|_{A \times B}\).
In consequence, by Equation~(\ref{definition:isolation:index}),
we have $D(a|B) + D(b|A) - ab= \alpha_S$ for all \(a \in A\), \(b \in B\) and,
so, choosing any  \(a \in A\) and \(b \in B\), we also have
\begin{eqnarray*}
\alpha_S &=&  D(a|B) + D(b|A) - ab\\
&=&  \frac{1}{2} \min_{a',a''\in A; b',b'' \in B}\{f(b') +f(b'') - b'b'' + f(a') +f(a'') -a'a'' \}\\
&=& D(f|B) + D(f|A)>0 \ \text{, as required}.
\end{eqnarray*}

Conversely, choosing arbitrary elements \(a_0 \in A\) and \(b_0 \in B\),
if $a_0b_0+a'b'=a_0b'+a'b_0$ holds for all $a'\in A$ and $b'\in B$ and the isolation index of $S$ is positive,
then, in view of Lemma~\ref{lemma:bivariate:map},  
we may choose any two non-negative real numbers $\gamma_A,\gamma_B$ with $\gamma_A+\gamma_B=\alpha_S$ and 
define the map 
\begin{equation}
\label{equation:split:map}
f=f_{A\ra\gamma_A,B\ra\gamma_B}:X\ra \R
\end{equation} 
by $f(a):=D(a|B) - \gamma_A$ for all $a\in A$ and 
$f(b):=D(b|A) - \gamma_B$ for all 
$b\in B$. In view of Equation~(\ref{definition:isolation:index}), we clearly have
$f(a)+f(b)=D(a|B) + D(b|A) - \alpha_S =ab$ for all  
$a\in A$ and $b\in B$.
Moreover, we have $f(a)+f(a')\ge aa'$ for all  
 $a,a'\in A$ as
$f(a)+f(a')\ge  D(a|B) +D(a'|B) -  2\alpha_S 
 = ab + a'b - 2D(b|A)
= aa' + (ab + a'b - aa')
 - 2D(b|A) \ge aa'$ holds for all  
 $a,a'\in A$ and every $b\in B$ in view of the definition of $D(b|A)$ \big(indeed, $ab + a'b - aa'$ is one of the terms whose minimum, over all $a,a'\in A$, coincides with $2D(b|A)$\big), and we have  $f(a)+f(a')> aa'$ for all  
 $a,a'\in A$ if and only 
 $\gamma_A<\alpha_S$
 holds as  $\gamma_A=\alpha_S$ implies $f(a)+f(a')= aa'$ for all $a,a'\in A$ with $ab + a'b - aa'
 =2D(b|A)$. By symmetry, we have also $f(b)+f(b')\ge bb'$ for all  
 $b,b'\in B$ and $f(b)+f(b') > bb'$ if and only if
 $\gamma_B<\alpha_S$
 holds. Thus, $E_f$ is a subset of $\binom{A}{2} \cup \binom{B}{2}$, and it coincides with this set if and only if $ \gamma_A,  \gamma_B< \alpha_S$ holds. 
So, $A|B$ must indeed be a block split, as required.
\hfill\(\blacksquare\)\\
 
It is also worth noting that, for every block split \(S=A|B\), every 
$f\in P(D)$ with $E_f\subseteq \binom{A}{2} \cup \binom{B}{2}$ (or, 
equivalently, with $f(a)+f(b)=ab$ for all $a\in A$ and  $b \in B$) actually is of the 
form $f=f_{A\ra\gamma_A,B\ra\gamma_B}$ for some 
$\gamma_A,\gamma_B\ge 0$ with $\gamma_A+\gamma_B=\alpha_S$: Indeed, 
in view of Equation~(\ref{definition:isolation:index}),
we have $D(a|B)-f(a)= ab +\alpha_S - D(b|A)  - f(a) = f(b) +\alpha_S - D(b|A)$ for 
all  $a\in A$ and $b\in B$ in this case, implying in particular that neither side 
changes once we replace $a$ by any other element in $A$ nor $b$ by any other element 
in $B$. So, choosing any fixed $a_0\in A$ and $b_0\in B$, we may put 
$\gamma_A:=D(a_0|B)-f(a_0)$ and $\gamma_B
 :=D(b_0|A)-f(b_0)$ in which case we have
 $\gamma_A+\gamma_B=D(a_0|B)+D(b_0|A)-f(a_0)-f(b_0)=D(a_0|B)+D(b_0|A)-a_0b_0=\alpha_S$, 
$f(a) = D(a|B)-\gamma_A$ and $f(b) = D(b|A)-\gamma_B$ for all $a\in A$ and  $b \in B$.
Moreover, we have $\gamma_A\ge 0$ in view of 
$D(a_0|B) = \frac{1}{2} \min\{a_0b+a_0b' - bb': b,b' \in B\}= f(a_0) +\frac{1}{2} \min\{f(b)+f(b') - bb': b,b' \in B\}\ge f(a_0)$,
and, similarly, \(\gamma_B \geq 0\). 

In other words, given any block split $S=A|B$, the set 
\[
T(D|S):=\{f\in P(D): E_f\subseteq \binom{A}{2} \cup \binom{B}{2}\}
\]
forms an straight line segment in \(\mathbb{R}^X\) parametrized by the straight line segment
$\{(\gamma_A, \gamma_B) \in \R_{\ge 0}^2: \gamma_A+ \gamma_B = \alpha_S\}$ in $\R^2$, and the two end points 
$f_A:= f_{A\ra\alpha_S ,B\ra 0}$ (closer to $A$) and 
$f_B:= f_{A\ra 0,B\ra\alpha_S }$ (closer to $B$) must each be either cut points of $T(D)$ that do not have a
neighborhood that is homeomorphic to the open interval 
\((-1,+1)\) or elements of the set \(K(D):=\{k_x:x\in X\}\) consisting of all Kuratowski maps 
for \(D\). Hence we have the following.

\begin{cor}
\label{corollary:endpoints:bridges}
For every block split \(S=A|B\) the maps 
\(f_A\) and \(f_B\) must be contained in the set 
\(Cut^*(D):= cut^*(D) \cup K(D)\). 
\end{cor}

%%%%%%%%%%%%%%%%%%%%%%%%%%%%%%%%%%%%%%%%%%%%%%%%%%%%%%%%%%%%%%%%%%%%%%%%%%%%%
\section{Key properties of $\Sigma_D$ and $Cut^*(D)$}
\label{section:key:properties}
%%%%%%%%%%%%%%%%%%%%%%%%%%%%%%%%%%%%%%%%%%%%%%%%%%%%%%%%%%%%%%%%%%%%%%%%%%%%%

As we have seen in the previous section, it is sometimes helpful to consider
the bigger set \(Cut^*(D)\) rather than \(cut^*(D)\). Since we can easily identify
those Kuratowski maps that are not in \(cut^*(D)\), 
we will now focus mainly on \(Cut^*(D)\). The following lemma establishes the 
key properties of $\Sigma_D$ and $Cut^*(D)$ that we will use in our algorithm 
to compute these sets recursively.

\begin{lemma}
\label{lemma:recursive:approach}
Let \(x\) be an arbitrary element of \(X\). 
Define \(X' := X \setminus \{x\}\)
and let \(D'\) denote the restriction of \(D\) to \(X'\).
Then the following assertions hold.
\begin{itemize}
\item[(i)]
If \(S=A|B\) is a block split of \(X\), then either \(S=\{x\}|X'\) or 
the restriction \(A \cap X'|B \cap X'\) of \(S\) to \(X'\) is a block split of \(X'\).
\item[(ii)]
If \(f \in Cut^*(D) \setminus K(D)=cut^*(D) \setminus K(D)\) has the property that there is no
block split \(S = A|B\) of \(X\) with \(f  \in \{f_A,f_B\}\),
then the restriction \(f'\) of \(f\) to \(X'\) is an element
of \(Cut^*(D')\) and \(f(x) = \max\{xy-f'(y) : y \in X'\}\) holds.
\end{itemize}
\end{lemma}

\noindent\textsl{Proof}:
(i) Clearly, if \(S=A|B\) is a block split of \(X\) with \(A,B \neq \{x\}\), then \(S' = A \cap X'| B \cap X'\) 
is a split of \(X'\), and there exists 
a map \(f \in P(D)\) such that \(\Gamma_f\) is the disjoint union of two cliques with
vertex sets \(A\) and \(B\) implying that the restriction \(f'\) of  \(f\) to \(X'\) is in \(P(D')\) and that \(\Gamma_{f'}\)
is the disjoint union of two cliques with vertex sets \(A\cap X'\) and \(B\cap X'\), respectively.
This establishes (i).

To see that (ii) holds, suppose \(f \in cut^*(D) \setminus K(D)\) and that there is no
block split \(S = A|B\) of \(X\) with \(f \in \{f_A,f_B\}\). 
Clearly, the restriction \(f'\) of \(f\) to \(X'\) is in \(P(D')\).
So, it remains to show that \(\Gamma_{f'}\) is disconnected, but 
not the disjoint union of two cliques, which implies in particular
that \(f' \in T(D')\).

Assume for a contradiction that \(\Gamma_{f'}\) is connected or
the disjoint union of two cliques. We first note that this implies that
\(\Gamma_f\) has at least one connected component that is a clique.
To see this, observe that if \(\Gamma_{f'}\) is connected,
then \(\Gamma_f\) has precisely two connected components, one of whom
consists of the single vertex \(x\), thus trivially forming a clique.
Similarly, if \(\Gamma_{f'}\) is the disjoint union of two cliques,
then one of these cliques is also a connected component of \(\Gamma_{f}\).

Let \(A\) denote the vertex set of a connected component of \(\Gamma_f\)
that forms a clique. Note that this implies that \(D(f|A)>0\).
Put \(B:=X \setminus A\) and \(S:=A|B\). 
We want to show that \(S\) is a block split with \(f \in \{f_A,f_B\}\), yielding
the required contradiction. To this end, choose arbitrary elements \(a_0 \in A\) 
and \(b_0 \in B\). Since \(f \in cut^*(D)\), \(B\) cannot be the vertex set of a
clique in \(\Gamma_f\), and so there must exist two distinct elements
\(b_1,b_2 \in B\) with the property that \(f(b_1)+f(b_2)=b_1b_2\), implying that
\(D(f|B) = 0\) holds. Since
\[a'b + ab' = f(a') + f(b) + f(a) + f(b') = a'b' + ab\]
clearly holds for all \(a,a' \in A\) and \(b,b' \in B\),
we have, in  view of Equation~(\ref{definition:isolation:index})
and the definition of \(\Gamma_f\), 
\begin{align*}
\alpha_S &= D(a_0|B) + D(b_0|A) - a_0b_0\\
         &= (f(a_0) + D(f|B)) + (f(b_0) + D(f|A)) - f(a_0) - f(b_0)\\
         &= D(f|A) > 0,
\end{align*}
and, therefore, \(S\) is indeed a block split.

It remains to show that \(f \in \{f_A,f_B\}\). More specifically, we will show
that \(f=f_B\) holds. By the definition
of \(f_B\) and in view of the fact that \(D(f|B)=0\) and
\(D(f|A) = \alpha_S\) holds, we have indeed
\(f_B(a) = D(a|B) = f(a) + D(f|B) = f(a)\),
for every $a\in A$, and
\(f_B(b) = D(b|A) - \alpha_S = f(b) + D(f|A) - \alpha_S = f(b)\),
for every $b\in B$, as claimed.
\hfill\(\blacksquare\)\\

We close this section with establishing bounds on the size of the sets 
\(\Sigma_D\) and \(Cut^*(D)\) that we will use in the analysis
of the run time of our algorithm in Section~\ref{section:algorith:computing:cut:d}. 

\begin{lemma}
\label{lemma:number:cut:points}
Let \(D\) be a metric on a finite set \(X\) with \(n\) elements.
Then  \(|\Sigma_D| \le 2n-3\) and \(|Cut^*(D)| \le 4n - 5\) holds.
\end{lemma}

\noindent\textsl{Proof}:
To establish the first claim, it suffices to
note that any two splits \(A_1|B_1, A_2|B_2 \in \Sigma_D\) 
are \emph{compatible}, that is, at least one 
of the four intersections \(A_1 \cap A_2\), \(A_1 \cap B_2\), 
\(B_1 \cap A_2\) and \(B_1 \cap B_2\) is empty, since it
is well known that every set of pairwise compatible splits of
\(X\) contains at most \(2n-3\) splits (see e.g. Proposition~2.1.3 and 
Theorem~3.1.4 in \cite{sem-ste-03a}). 
So, assume for a contradiction that there exist two splits \(A_1|B_1\)
and \(A_2|B_2\) in \(\Sigma_D\) that are not compatible. Then we can
choose arbitrary elements \(a \in A_1 \cap A_2\), \(b \in B_1 \cap A_2\),
\(c \in A_1 \cap B_2\) and \(d \in B_1 \cap B_2\). By the definition
of a block split, there exist maps \(f_i \in T(D)\), \(i \in \{1,2\}\), 
for which the graph \(\Gamma_{f_i}\) is the 
disjoint union of two cliques with vertex sets \(A_i\) and \(B_i\).
But then, by the definition of \(\Gamma_{f_1}\) and \(\Gamma_{f_2}\),
\begin{align*}
f_1(a)+f_1(b)+f_1(c)+f_1(d) &= ab + cd < f_2(a)+f_2(b)+f_2(c)+f_2(d)\\
&= ac+bd < f_1(a)+f_1(b)+f_1(c)+f_1(d)
\end{align*}
holds, a contradiction. 

Next we show \(|Cut^*(D)| \le 4n - 5\). Since, clearly, \(|K(D)| \le n\), it
suffices to show that \(|Cut^*(D) \setminus K(D)| \le 3n - 5\). In \cite{dre-hub-koo-08a},
it is shown that there exists a block realization \(G=G_D\) of \(D\) such that
the cut vertices in \(G\) are in one-to-one correspondence with the elements
in \(Cut^*(D) \setminus K(D)\). Moreover, the number of cut vertices in any graph
is well known to be less than the number of blocks of this graph (see e.g. \cite{har-pri-66a}).
Hence, it suffices to show that the number of blocks in \(G\) is at most $3n-5$. 
Yet, it has been shown in \cite{dress:huber:compatible:decompositions:2008} that
there is a canonical bijection from the set of blocks of \(G\) to a set \(\Pi\) of
\emph{$($\!strongly$)$ compatible} partitions of \(X\), that is, of 
partitions such that there exist, for any two distinct partitions 
$\pi_1$ and $\pi_2$, two necessarily unique subsets \(A_1 \in \pi_1\) and \(A_2 \in \pi_2\) of $X$ with 
\(A_1 \cup A_2 = X\) (generalizing the concept of compatibility
for splits to arbitrary partitions of \(X\)). 
Therefore, it suffices to show that, for all $n\ge 2$, every set of pairwise compatible partitions of \(X\)
contains at most \(3n-5\) partitions which we will establish 
by induction on the size of \(X\). Clearly, if \(n=2\) then there is only one partition of \(X\).

Now assume \(n=|X| > 2\). If every partition in \(\Pi\) is
a split of \(X\), then \(|\Pi| \leq 2n-3 < 3n-5\) must hold.
Otherwise, there exists a partition \(\pi \in \Pi\) that
contains at least three subsets of \(X\). For every \(A \in \pi\), fix
an arbitrary element \(x_A \in X \setminus A\), define
\(\Pi_A\) to be the set of the restrictions \(\pi'_{|A \cup \{x_A\}}\) of those partitions  
\(\pi' \in \Pi\) with the property that there exists some \(A' \in \pi'\) with \(A \cup A' = X\), and note that any such partition $\pi'$ can be recovered from its restriction \(\pi'_{|A \cup \{x_A\}}\) as it must consist of all subsets $B$ in that restriction that do not contain $x_A$ and the complement of their union. Thus, it is not hard to see that, for every \(A \in \pi\), any
two partitions of \(A \cup \{x_A\}\) in \(\Pi_A\) are compatible,
that \(1 + \sum_{A \in \pi} |\Pi_A| = |\Pi|\) holds, and that \(|A \cup \{x_A\}| < |X|\) holds for every \(A \in \pi\). Hence, by induction,
\[
|\Pi| = 1 + \sum_{A \in \pi} |\Pi_A| \leq 1 + \sum_{A \in \pi} (3|A| - 2) \leq 3n-5,
\] 
as required.
\hfill\(\blacksquare\)

%%%%%%%%%%%%%%%%%%%%%%%%%%%%%%%%%%%%%%%%%%%%%%%%%%%%%%%%%%%%%%%%%%%%%%%%%%%%%
\section{The algorithm for computing $Cut^*(D)$}
\label{section:algorith:computing:cut:d}
%%%%%%%%%%%%%%%%%%%%%%%%%%%%%%%%%%%%%%%%%%%%%%%%%%%%%%%%%%%%%%%%%%%%%%%%%%%%%

In this section, we present our new algorithm for computing \(Cut^*(D)\)  
called \textsc{ComputeCutPoints}(\(D\)) which follows the recursive
approach suggested by Lemma~\ref{lemma:recursive:approach}.
This algorithm can be regarded as a speed-up of the algorithm for computing
cutpoints presented in \cite{dress:virtual:cutpoints:2007}, 
which, as mentioned in the introduction, also improves upon the
run time of the algorithm presented in \cite{hertz:varone:cutpoint:partition}.
In Figure~\ref{algorithm:cut:points}, we present a pseudocode
for this algorithm. Besides \(Cut^*(D)\) the algorithm returns
the set \(\Sigma_D\) and the auxiliary set \(\mathcal{A}(\Sigma_D)\),
which, for every split \(S=A|B \in \Sigma_D\), contains
the 4-tuple \((a_S,b_S,D(a_S|B),D(b_S|A))\), where
\(a_S \in A\) and \(b_S \in B\) are fixed elements that are
arbitrarily chosen during the course of the algorithm.

To illustrate how our algorithm 
computes \(Cut^*(D)\), consider the metric \(D\) presented in
Figure~\ref{figure:block:realization}(a). Suppose in Line~3
of the pseudocode in Figure~\ref{algorithm:cut:points}, 
we select the element \(c\). Consider
the restriction \(D'\) of \(D\) to the subset \(X' := X \setminus \{c\}\).
A block realization of \(D'\) is presented in 
Figure~\ref{figure:block:realization}(c). It is easy to check
that the set of block splits of \(D'\) is
\[
\Sigma' = \{\{a\}|\{b,d,e\}, \{b\}|\{a,d,e\}, \{d\}|\{a,b,e\}, \{a,b\}|\{d,e\}\}.
\]
Note that the splits in \(\Sigma'\) are in one-to-one correspondence
with the edges of the block realization in Figure~\ref{figure:block:realization}(c).
The set \(C' := Cut^*(D')\) consists of the Kuratowski maps in \(K(D')\) and
one additional map \(f \in \mathbb{R}^{X'}\) with
\(f(a)=2\), \(f(b)=1\), \(f(d)=7\) and \(f(e)=3\).
Note that this map corresponds to the cut vertex \(v\)
in Figure~\ref{figure:block:realization}(c) as \(f(x)\)
equals the length of a shortest path from \(v\) to
\(x\) in the block realization for every \(x \in X'\).

Given \(C'\) and \(\Sigma'\), the algorithm first computes
the set \(\Sigma\) of block splits of \(D\) and
the auxiliary set \(\mathcal{A}\) (Lines~6-21).
In our example it is easy to check that each of the splits
in \(\Sigma'\) gives rise to precisely one split in 
\(\Sigma\), that is,
\[
\Sigma = \{\{a\}|\{b,c,d,e\}, \{b\}|\{a,c,d,e\}, \{d\}|\{a,b,c,e\}, \{a,b\}|\{c,d,e\}\}.
\]
Next the set \(C := Cut^*(D)\) is computed (Lines~22-27) by first
adding the maps \(f_A\) and \(f_B\) for every \(S =A|B \in \Sigma\).
For the metric \(D\) in Figure~\ref{figure:block:realization}(a), this
yields, in addition to the Kuratowski maps \(k_a\), \(k_b\) and \(k_d\),
the 3 cutpoints
\((2,1,4,7,3)\), \((3,2,3,6,2)\) and \((8,7,2,1,3)\),
where \((x_1,x_2,\dots,x_5) \in \mathbb{R}^5\) represents the
map \(f \in \mathbb{R}^X\) with \((x_1,x_2,\dots,x_5) = (f(a),f(b),\dots,f(e))\).
Note that these cutpoints correspond to the 3 cut vertices in the block realization
of \(D\) in Figure~\ref{figure:block:realization}(b).
For our example, the computation of \(C\) is completed
by adding the Kuratowski maps \(k_c\) and \(k_e\) (Line~27).

\begin{figure}
\begin{tabbing}
XXX\= XX\= XX\=  XX\= XX\= XX\= XX\=  XXXXX\=  XXXXXXXXXXX\= \kill \\
{\large \textsc{ComputeCutPoints}($D$)}\\
\rule{13cm}{0.5pt}\\
Input: \> \> \>a metric $D$ on $X$\\
Output: \> \> \>$Cut^*(D)$, \(\Sigma_D\), \(\mathcal{A}(\Sigma_D)\)\\
\rule{13cm}{0.5pt}\\
1. \> \textbf{if} \(X=\{x\}\), \textbf{then} \textbf{return} \(C:=\{k_x\}\),
      \(\Sigma := \emptyset\) and \(\mathcal{A}:=\emptyset\).\\
2. \> Initialize \(C := \emptyset\), \(\Sigma := \emptyset\) and \(\mathcal{A}:=\emptyset\).\\
3. \> Select \(x \in X\) arbitrarily.\\
4. \> Put \(X' := X \setminus \{x\}\), and let \(D'\) denote the restriction of \(D\) to \(X'\).\\
5. \> Compute recursively \(C' := Cut^*(D')\), \(\Sigma':=\Sigma_{D'}\) and \(\mathcal{A}' := \mathcal{A}(\Sigma_{D'})\).\\
6. \> \textbf{for each} \(S'=A'|B' \in \Sigma'\) \textbf{do}\\
7. \> \> Put \(a_S := a_{S'}\) and \(b_S := b_{S'}\).\\
8. \> \> Put \(A:= A' \cup \{x\}\), \(B := B'\) and extend \(S'\) to \(S := A|B\).\\
9. \> \> Compute \(D(a_S|B) := D(a_{S'}|B')\).\\
10. \> \> Compute \(D(b_S|A) := \min \{D(b_{S'}|A'), \frac{1}{2} \underset{a \in A}{\min} \ (b_Sx+b_Sa-ax) \}\).\\
11. \> \> \textbf{if} \(S\) is a block split of \(X\), \textbf{then}\\
12. \> \> \> Insert \(S\) into \(\Sigma\) and \((a_S,b_S,D(a_S|B),D(b_S|A))\) into \(\mathcal{A}\).\\
13. \> \> Put \(A:= A'\), \(B := B' \cup \{x\}\) and extend \(S'\) to \(S := A|B\).\\
14. \> \> Compute \(D(a_S|B) := \min \{D(a_{S'}|B'), \frac{1}{2} \underset{b \in B}{\min} \ (a_Sx+a_Sb-ax)\}\).\\
15. \> \> Compute \(D(b_S|A) := D(b_{S'}|A')\).\\
16. \> \> \textbf{if} \(S\) is a block split of \(X\), \textbf{then}\\ 
17. \> \> \> Insert \(S\) into \(\Sigma\) and \((a_S,b_S,D(a_S|B),D(b_S|A))\) into \(\mathcal{A}\).\\
18. \> Put \(S = \{x\}|X'\), \(a_S := x\) and select \(b_S \in X'\) arbitrarily.\\
19. \> Compute \(D(a_S|X')\) and \(D(b_S|\{x\})\).\\
20. \> \textbf{if} \(S\) is a block split of \(X\), \textbf{then}\\
21. \> \> Insert \(S\) into \(\Sigma\) and \((a_S,b_S,D(a_S|X'),D(b_S|\{x\}))\) into \(\mathcal{A}\).\\ 
22. \> \textbf{for each} \(S = A|B \in \Sigma\) \textbf{do}\\
23. \> \> Insert \(f_A\) and \(f_B\) into \(C\).\\
24. \> \textbf{for each} \(f' \in C'\) \textbf{do}\\
25. \> \> Extend \(f'\) to \(f \in \mathbb{R}^X\) putting \(f(x) := \max\{xy-f'(y) : y \in X'\}\).\\
26. \> \> \textbf{if} \(f\) is a cutpoint of \(D\), 
         \textbf{then} insert \(f\) into \(C\).\\
27. \> \textbf{for each} \(x \in X\) \textbf{do} insert \(k_x\) into \(C\).\\
28. \> \textbf{return} \(C\), \(\Sigma\) and \(\mathcal{A}\).
\end{tabbing}
\caption{Pseudocode for our algorithm for computing \(Cut^*(D)\).}
\label{algorithm:cut:points}
\end{figure}

\begin{theorem}
\label{theorem:algorithm:cut:points}
Given a metric \(D\) on a set \(X\) with \(n\) elements, the
algorithm \textsc{ComputeCutPoints}(\(D\)) computes \(Cut^*(D)\)
in \(O(n^3)\) time.
\end{theorem}

\noindent\textsl{Proof}:
We first show that our algorithm is correct. To do this we use induction
on the size \(n\) of \(X\). Our induction hypothesis is that our
algorithm computes \(Cut^*(D)\) and the set \(\Sigma_D\) of block splits of \(X\) correctly.
If \(|X| = 1\), there is nothing to prove.
Now suppose that \(|X| > 1\) holds. 
Let \(x\) be the element in \(X\) selected by our algorithm (Line~3), 
put \(X' := X \setminus \{x\}\), and let \(D'\) denote
the restriction of \(D\) to \(X'\) (Line~4).
By Lemma \ref{lemma:recursive:approach}(i), the set \(\Sigma_D\) of block splits of \(X\)
can be computed from the set \(\Sigma_{D'}\) of block splits of \(X'\).
By induction, the recursive call (Line~5) will correctly compute \(\Sigma_{D'}\) and,
therefore, our algorithm will correctly compute \(\Sigma_D\) (Lines~6-21).
Similarly, by Corollary~\ref{corollary:endpoints:bridges}
and Lemma~\ref{lemma:recursive:approach}(ii), the set \(Cut^*(D)\) can be computed
from \(\Sigma_D\) and \(Cut^*(D')\). We have argued already that the computation
of \(\Sigma_D\) is correct and, again by induction, the recursive call (Line~5) 
will correctly compute \(Cut^*(D')\). Hence, our algorithm will correctly 
compute \(Cut^*(D)\) (Lines~22-27). 

We next show that our algorithm has run time \(O(n^3)\).
We claim that an upper bound \(T(n)\) on the run time
will satisfy the recurrence \(T(n) \leq T(n-1) + O(n^2)\).
Using standard techniques for solving recurrences 
(see e.g. \cite{cormen:intro:algorithms:2001}),
this yields \(T(n) \in O(n^3)\). So, it remains to show
that all operations except those performed in the recursive call (Line~5)
can be done in \(O(n^2)\) time. 

We first focus on the computation of \(\Sigma_D\) from \(\Sigma_{D'}\) (Lines~6-21).
Let \(S'=A'|B'\) be an arbitrary split in \(\Sigma_{D'}\). We can assume
that \(D(a_{S'}|B')\) and \(D(b_{S'}|A')\) are available 
from the 4-tuple \((a_{S'},b_{S'},D(a_{S'}|B'),D(b_{S'}|A')) \in \mathcal{A}'\).
We want to check whether the split 
\(S = A|B = A' \cup \{x\}|B'\) is a block split of \(X\) (Line~11).
By Lemma \ref{lemma:characterization:block:splits} it suffices to check
whether \(\alpha_S>0\)
and \(a_Sb + ab_S = a_Sb_S + ab\) holds for all \(a \in A\), \(b \in B\),
using \(a_S = a_{S'}\) and \(b_S = b_{S'}\).
Note that, since \(S'\) is a block split of \(X'\), it suffices to
check whether \(a_Sb + xb_S = a_Sb_S + xb\) holds for all \(b \in B\), which can
be done in \(O(n)\) time. 
Moreover, since 
\(D(a_S|B) = D(a_{S'}|B')\) and
\[
D(b_S|A) = \min \{D(b_{S'}|A'), 
                             \frac{1}{2} \min \{b_Sx+b_Sa-ax:a \in A' \cup \{x\}\}\}
\]
hold (Lines~9-10), we can also compute
\(\alpha_S = D(a_S|B) + D(b_S|A) - a_Sb_S\) in \(O(n)\) time. 

To summarize, whether \(S\) is a block split of \(X\) or not
can be checked in \(O(n)\) time. Using completely similar arguments, it can
also be shown that checking whether \(A'|B' \cup \{x\}\) is a block split of \(X\) (Line~16)
can be done in \(O(n)\) time. Note that, by Lemma~\ref{lemma:number:cut:points}, 
there are \(O(n)\) block splits of \(D'\).
Thus, our algorithm will perform \(O(n)\) iterations of the loop in Line 6 
and each iteration is completed in \(O(n)\) time, yielding \(O(n^2)\) in total
for Lines~6-17.

To finish the computation of \(\Sigma_D\), we need to check whether
the split \(S = \{x\}|X'\) is a block split of \(X\) (Lines~18-21). 
To do this, we fix \(a_S = x\) and choose an arbitrary \(b_S \in X'\).
Then, we compute \(D(a_S|X')\) and \(D(b_S|\{x\})\),
which can be done in \(O(n^2)\) time, and check whether 
\(\alpha_S = D(a_S|X') + D(b_S|\{x\}) - a_Sb_S > 0\) holds.
We also check whether \(a_Sb + xb_S = a_Sb_S + xb\) holds for all \(b \in X'\),
which can be done in \(O(n)\) time. 
This finishes the analysis of the time needed to compute \(\Sigma_D\).

Next, we focus on the computation of \(Cut^*(D)\) (Lines~22-27).
We use a data structure \textsc{Dic}
to store the elements in \(Cut^*(D)\) computed so far. 
Since, by Lemma \ref{lemma:number:cut:points},
\(|Cut^*(D)| \in O(n)\), the data structure \textsc{Dic} can be implemented in such a way
that inserting a single element of \(Cut^*(D)\) into \textsc{Dic} and, later on, 
checking whether an element of \(Cut^*(D)\) has
already been stored in \textsc{Dic} both takes \(O(n)\) time, see e.g. \cite{gon-00a}.
Moreover, we assume that, for every \(f' \in Cut^*(D')\),
the connected components of the graph \(\Gamma_{f'}\) have been computed
and the cliques among them have been marked. 

So, first consider an arbitrary block split \(S = A|B \in \Sigma_D\).
If we have \(A=\{x\}\) and \(B=X'\), then
we compute \(f_Y\) along with the connected components of \(\Gamma_{f_Y}\),
marking the cliques among them, in \(O(n^2)\) time for all \(Y \in \{A,B\}\). 
Next we consider the case that there exists some \(S' = A'|B' \in \Sigma_{D'}\)
such that \(A = A' \cup \{x\}\) and \(B = B'\) (the following argument is completely 
analogous if \(A = A'\) and \(B = B' \cup \{x\}\)).
Let \(a_S \in A'\) and \(b_S \in B'\) be the elements that we fixed
for \(S\) in the course of the algorithm and
let \(f_{A'}\) and \(f_{B'}\) be the maps in \(Cut^*(D')\)
associated with the split \(S'\).
Then we have 
\[f_B(a) = D(a|B) = D(a_S|B) - a_Sb_S + ab_S = D(a_S|B') - a_Sb_S + ab_S = f_{B'}(a)\] 
for all \(a \in A'\) and
\[f_B(b) = D(b|A) - \alpha_S = a_Sb - D(a_S|B) = a_Sb - D(a_S|B') = f_{B'}(b)\]
for all \(b \in B = B'\), since \(D(a_S|B) = D(a_S|B')\) clearly holds.
Hence, computing \(f_B\), the connected components of \(\Gamma_{f_B}\) 
and marking the cliques among them can be done in \(O(n)\) time, based on 
\(f_{B'}\) and the connected components of \(\Gamma_{f_{B'}}\).
Similarly, if \(D(b_S|A) = D(b_S|A')\) holds,
\(f_A\), the connected components of \(\Gamma_{f_A}\)
and the cliques among them can be computed in \(O(n)\) time.
Otherwise, that is, if \(D(b_S|A) < D(b_S|A')\) holds,
the graph induced by \(\Gamma_{f_A}\)
on \(X'\) is the disjoint union of two cliques
with vertex sets \(A'\) and \(B'\), respectively. 
To see this, note that
\(f_A(a) + f_A(a')  > f_{A'}(a) + f_{A'}(a') \geq aa'\),
\begin{align*}
f_A(b) + f_A(b')  &= 2\alpha_S + a_Sb + a_Sb' - 2 D(a_S|B) \\
                  &> a_Sb + a_Sb' - 2 D(a_S|B') = f_{B'}(b) + f_{B'}(b') \geq bb',
\end{align*} 
and \(f_A(a) + f_A(b) = ab = f_{A'}(a) + f_{A'}(b)\)
holds for all \(a,a' \in A'\) and \(b,b' \in B'\).
But then, also in this case, the connected components of \(\Gamma_{f_A}\)
and the cliques among them can easily be computed in \(O(n)\) time.

It remains to consider an arbitrary \(f' \in Cut^*(D')\) (Lines 24-26).
Extending \(f'\) to \(f\) (Line 25), that is, computing
\(f(x)\) can be done in \(O(n)\) time.
Recall that we assume that the connected components of the 
graph \(\Gamma_{f'}\) and the cliques among them have been computed.
From this information, we can compute in \(O(n)\) time
the connected components of \(\Gamma_f\) and determine which
of them are cliques. Hence the loop in Line~24 will take \(O(n^2)\) time,
as required. Similarly, the loop in Line~27 will also take \(O(n^2)\) time.
This finishes the analysis of the run time of our algorithm and thus the proof of
the theorem.
\hfill\(\blacksquare\)\\

\subsubsection*{Acknowledgments}
Authors Moulton and Spillner were supported by the 
Engineering and Physical Sciences Research Council 
[grant number EP/D068800/1].  
A.\,Dress thanks the Chinese Academy of Sciences, 
the Max-Planck-Gesellschaft, and the German BMBF for
their support, as well as the Warwick Institute for Advanced Study
where, during two wonderful weeks, the basic outline of this paper was
conceived. Huber and Koolen thank the Royal Society for their
support in the context of a International Joint Project
grant. Koolen was also partially supported by the Korea 
Research Foundation of the Korean Government under grant 
number KRF-2007-412-J02302. 
We would also like to thank the anonymous referees
for their helpful comments on earlier versions of
this paper.

%\bibliographystyle{plain}
%\bibliography{cutpoints} 

\end{document}